\documentclass[sigconf, nonacm]{acmart}

\usepackage{tikz}
\usetikzlibrary{positioning,fit}
\usepackage{array}
\usepackage{booktabs}
\usepackage{algorithm}
\usepackage[noend]{algpseudocode}
\usepackage{graphicx}
\usepackage{xcolor}
\usepackage{colortbl}
\usepackage{xspace}
\usepackage{makecell}
\usepackage{amsmath}
\usepackage{multirow}
\usepackage{subcaption}
\usepackage{dsfont}
\definecolor{light-gray}{gray}{0.90}

\newcommand{\model}{\textsc{Octopus}\xspace}

\usepackage[most]{tcolorbox}
\usepackage{pifont}
\newcommand{\cmark}{\ding{51}} 
\newcommand{\xmark}{\ding{55}}

\usepackage{enumitem}

\newtcolorbox{graybox}[1][]{
  colback=gray!10,
  colframe=black!70,
  boxrule=0.5pt,
  arc=2pt,
  left=6pt, right=6pt, top=6pt, bottom=6pt,
  fonttitle=\bfseries,
  title=#1
}

\newtheorem{definition}{Definition}[section]

\newtheorem{example}{Example}[section]

\usepackage{pvldb}

\renewcommand\vldbdoi{XX.XX/XXX.XX}
\renewcommand\vldbpages{XXX-XXX}
\renewcommand\vldbavailabilityurl{https://github.com/wenzhilics/octopus} 

\begin{document}
\title{\model: Agentic Multi-Table Discovery over Tabular Data Lakes}

\author{Wen-Zhi Li}
\affiliation{%
  \institution{Cornell University}
  \city{Ithaca}
  \country{USA}}
\email{wenzhi@cs.cornell.edu}

\author{Sainyam Galhotra}
\affiliation{%
  \institution{Cornell University}
  \city{Ithaca}
  \country{USA}}
\email{sg@cs.cornell.edu}

\begin{abstract}
Tabular data discovery aims to identify the tables relevant to a natural-language question from a large collection of heterogeneous data. Existing systems typically assume that a question is answered by a single table and rely on retrieval pipelines constructed through substantial offline preprocessing. These assumptions become restrictive when relevant information is distributed across multiple tables. In this paper, we present \model, a lightweight agentic framework for multi-table data discovery. \model follows a thin-harness design that combines lightweight offline semantic guidance with adaptive access to raw tabular data. At query time, an LLM agent dynamically orchestrates semantic column search, structured tabular operations, and general-purpose execution to locate and verify relevant tables based on intermediate observations. We further introduce a benchmark that explicitly evaluates multi-table data discovery, covering both settings where multiple tables independently answer a question and settings where multiple tables must be joined. Experimental results across four independent-discovery datasets and two join-based datasets show that \model achieves the highest macro-F1 on all six datasets. It also substantially reduces offline preprocessing overhead relative to Pneuma and online latency and API cost relative to Claude Code. When coupled with the same NL2SQL agent, \model yields substantially higher downstream question-answering F1 than Claude Code, showing that improved table discovery provides more complete evidence for downstream analytics.
Code is available at \url{https://github.com/wenzhilics/octopus}.
\end{abstract}

\maketitle


\section{Introduction}

\begin{table*}[ht]
\centering
\caption{Comparison of the problem settings supported by existing data discovery systems, NL2SQL methods, and our system \model. Data Source indicates whether a method is designed to operate over a curated database or a large-scale heterogeneous data lake. Ground Truth Tables indicates whether the target consists of a single table, multiple tables that must be joined to answer the query, or multiple tables that can independently answer the query.}
\label{table:model_compare}
\begin{tabular}{cc|cc|ccc}
\toprule
\multicolumn{2}{c|}{\multirow{2}{*}{\textbf{Methods}}}
  & \multicolumn{2}{c|}{Data Source}
  & \multicolumn{3}{c}{Ground Truth Tables} \\
\cmidrule(lr){3-4}\cmidrule(lr){5-7}
& & Curated Database & Data Lake & Single & Multi. (requires join) & Multi. (independent) \\
\midrule
\multicolumn{2}{l|}{Data Discovery Systems~\cite{opendtr,gtr,solo,pneuma}}
& \cmark & \cmark & \cmark & \xmark & \xmark \\

\multicolumn{2}{l|}{NL2SQL~\cite{text2sql1,text2sql2,text2sql3,text2sql4}}
& \cmark & \xmark & \cmark & \cmark & \xmark \\

\multicolumn{2}{l|}{\model}
& \cmark & \cmark & \cmark & \cmark & \cmark \\
\bottomrule
\end{tabular}
\end{table*}

Tabular data constitute a fundamental form of information across large repositories such as data lakes, databases, and cloud storage systems~\cite{datalake,lakehouse,dataintegration}. Among the vast number of available tables, only a small subset is typically relevant to a given user intent. Identifying such relevant tables defines the data discovery problem~\cite{datadiscovery,gent,ver}, a critical preprocessing step for downstream data analytics and integration tasks.

Recent data discovery systems have introduced natural language (NL) interfaces that allow users to express intent in free-form queries~\cite{solo,pneuma}. However, these systems often assume that each question can be answered by a single table—an assumption that rarely holds in practice. In many cases, multiple tables can independently provide valid answers, and returning all of them may yield complementary insights. In others, a question can only be answered by combining information across multiple tables (e.g., via joins), where no individual table alone suffices. Ideally, a system should return all tables relevant to the question.
Subsequently, for specific downstream tasks, the user may expect the system to return analytical insights extracted from the discovered tables~\cite{spider}.
In this paper, unlike prior systems that assume a single-table setting, we study both multi-table independent and multi-table join scenarios, where answers may either emerge from several standalone tables or from combining complementary ones.
Table~\ref{table:model_compare} summarizes how our setting extends the scope of existing systems.

Most existing data discovery systems rely on time- and memory-intensive offline preparation, including costly training or indexing~\cite{opendtr,gtr,solo}, which can hinder deployment and transfer across domains. 
Recent advances in large language models (LLMs)~\cite{gpt4,llama,qwen} and retrieval-augmented generation (RAG)~\cite{rag} mitigate this limitation by leveraging LLMs to directly retrieve relevant information from natural-language queries, as exemplified by the state-of-the-art system Pneuma~\cite{pneuma}.
However, Pneuma indexes only a sampled subset of rows to reduce memory usage, which can fail when user queries refer to specific values absent from the sampled rows.
More fundamentally, RAG systems rely on a relatively static retrieval mechanism that matches queries against indexed table chunks using lexical or semantic retrievers. Such a mechanism cannot easily adapt to complex reasoning tasks, and chunks filtered out at an early retrieval stage cannot be recovered later.

Entering the agentic era~\cite{rlm,dci}, a new paradigm of retrieval agents offers a promising direction for lightweight data discovery. Instead of relying on a predefined retrieval pipeline, these agents can directly interact with the raw corpus through general-purpose tools, such as file reads, \texttt{grep}, and executable shell commands~\cite{claudecode,codex}. Powered by the capability of backbone LLMs, they can autonomously plan retrieval strategies and synthesize executable retrieval primitives on-the-fly. Also, by observing intermediate execution results, they can iteratively refine the plans and primitives to adapt dynamically to the query and the underlying data~\cite{react}.
The potential of such agentic interaction has already been demonstrated in domains such as software engineering~\cite{swebench,codex} and complex, long-horizon tasks in interactive terminal environments~\cite{terminalbench}. This capability opens up a new possibility for lightweight data discovery over raw tables, where retrieval strategies can be synthesized and adapted on demand rather than predefined in advance.

However, while agentic retrieval largely eliminates the heavy offline preparation required by conventional indexing pipelines and provides the flexibility to access arbitrary values, it can still be expensive at scale. In particular, repeatedly running \texttt{grep} or custom shell commands over a large data lake incurs substantial cost. This raises a natural question: \emph{Can a lightweight offline investment make agentic data discovery practical in terms of both accuracy and cost?}

Through our study, we find that column-header embeddings provide an effective and lightweight form of semantic guidance for agentic data discovery.
Given these precomputed embeddings, the agent can invoke a semantic column-search tool using arbitrary concepts or entities that it derives from the original question. This provides the agent with targeted semantic guidance and avoids inefficient blind search, where the agent must repeatedly revise keywords and rerun \texttt{grep} when earlier searches fail to match the desired values.

Building on these observations, we introduce \model, a minimalist agentic framework for tabular data discovery in both multi-table independent and multi-table join settings. \model adopts a ``thin harness'' design~\cite{thinharness}: rather than prescribing a fixed retrieval pipeline, it exposes complementary information-access mechanisms through a compact set of tools and delegates their orchestration to the agent.
These tools include column-level semantic search, domain-specific operators for exploring tabular data, and general-purpose primitives such as \texttt{grep} and executable shell commands. This design preserves the flexibility of agentic retrieval while providing the structured access needed to make large-scale data discovery both effective and efficient.

To comprehensively evaluate \model under this new setting, we introduce a new benchmark tailored to both multi-table independent and multi-table join discovery. Although several benchmarks have been developed for table discovery and related tabular reasoning tasks~\cite{opendtr,fetaqa,qatch,cmdbench,pneuma,spider,bird}, they do not jointly cover the two forms of multi-table discovery studied in this work: retrieving multiple independently sufficient tables and discovering a set of tables that must be joined to answer the question.
Our benchmark comprises four datasets for multi-table independent discovery, where multiple tables can independently answer a query; and two datasets for multi-table join discovery, where information from multiple tables must be combined to derive the answer. Across all datasets, \model consistently achieves the highest F1 scores while incurring substantially lower computational overhead. 
Moreover, when integrated with the same NL2SQL agent, the tables retrieved by \model yield substantially higher downstream question-answering F1 than those retrieved by Claude Code, demonstrating that its retrieval gains translate to downstream task performance.

We summarize the main contributions of this work as follows:

\begin{itemize}
    \item We extend the data discovery problem beyond the conventional single-table setting and formulate a unified multi-table discovery setting that covers both independent and join-based scenarios.

    \item We identify fine-grained, precomputed column-header embeddings as a particularly effective and lightweight form of offline preparation for agentic tabular data discovery. Building on this finding, we propose \model, a minimalist agentic framework that combines semantic column search, domain-specific data-access tools, and general-purpose terminal primitives under a thin-harness design.

    \item We construct a comprehensive benchmark for multi-table data discovery, comprising four datasets for multi-table independent discovery and two datasets for multi-table join discovery.

    \item Extensive experiments show that \model achieves the highest macro-F1 across all six datasets and the highest exact-match rate on both join-based datasets. \model further reduces offline preprocessing overhead relative to Pneuma and online latency and API cost relative to Claude Code. When coupled with the same NL2SQL agent, \model yields higher downstream question-answering F1 than Claude Code across all six datasets.
\end{itemize}

The remainder of this paper is organized as follows. We first introduce the preliminaries and the formal problem definition in Section~\ref{section:preliminaries}. We then introduce \model in Section~\ref{sec:octopus}. Next, we introduce how we generate the benchmark to evaluate \model in Section~\ref{section:table_discovery_benchmark}. Then, we provide empirical evaluation in Section~\ref{section:evaluation}. Finally, we discuss related work in Section~\ref{section:related_work} and conclude the paper in Section~\ref{section:conclusions}.

\section{Preliminaries}\label{section:preliminaries}

\begin{figure*}[!t]
\centering
\includegraphics[width=0.95\linewidth]{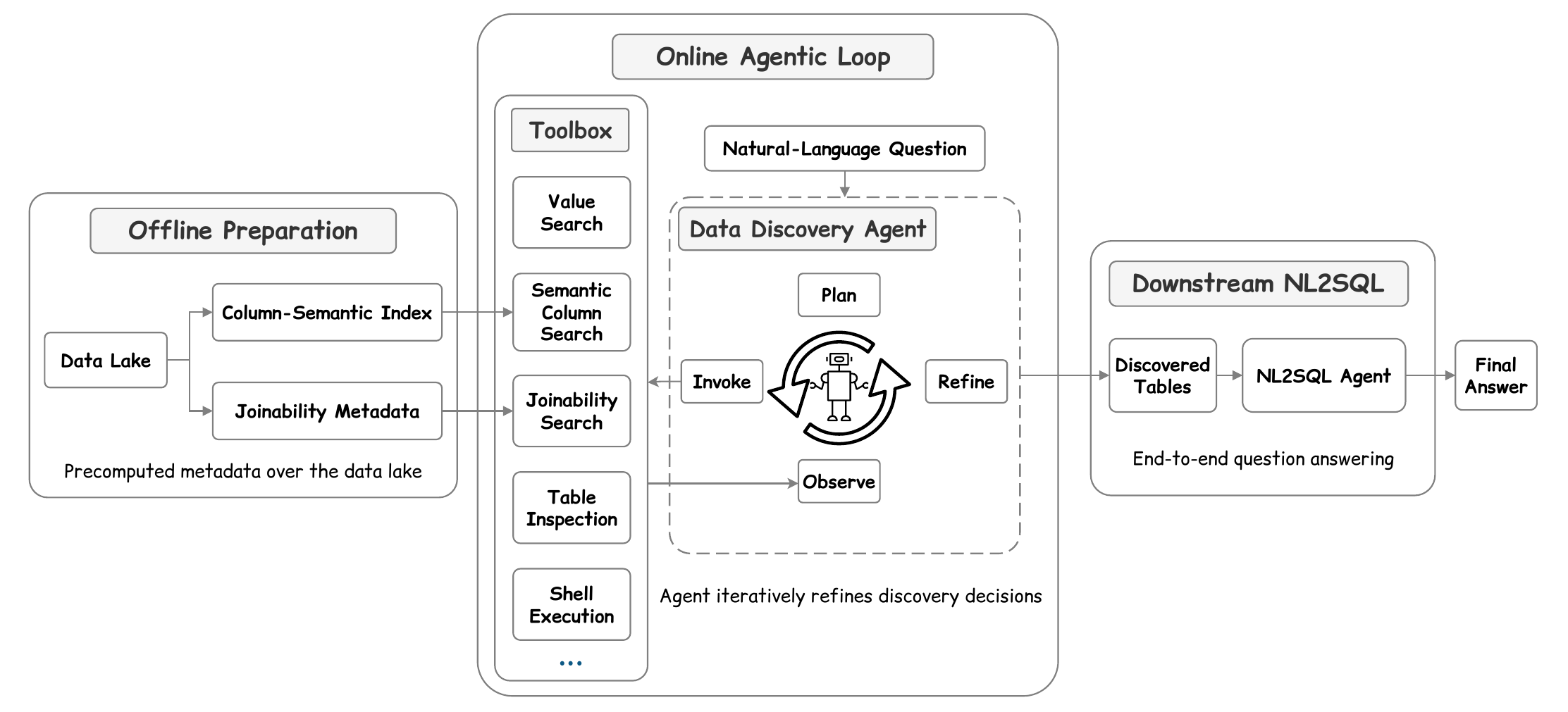}
\centering
\caption{System overview of \model. \model consists of a lightweight offline preparation stage that constructs reusable semantic and joinability metadata over the data lake, and an online agentic discovery stage that follows a thin-harness design, allowing the agent to iteratively plan, invoke tools, observe intermediate results, and refine its retrieval decisions. The discovered tables can further be passed to a downstream NL2SQL agent for end-to-end question answering.}\label{fig:octopus_overview}
\end{figure*}

In this section, we formalize different types of tabular data discovery tasks considered in this work and illustrate each setting with motivating examples.

\begin{definition}[Single-Table Discovery]
Let \( \mathcal{T} = \{ T_1, \ldots, T_n \} \) be a collection of \( n \) tables, and let \( Q \) be an NL question that is relevant to exactly one table \( T_Q \in \mathcal{T} \).  
A table discovery system \( S \) aims to identify this relevant table: $\hat{T}_Q = S(Q, \mathcal{T}) \approx T_Q$.

\end{definition}

\begin{example}
The question ``What is the population of New York City in 2020?'' can be answered by a single U.S.\ Census table containing city populations by year.
\end{example}

\begin{definition}[Multi-Table Independent Discovery]
Let \( \mathcal{T} = \{ T_1, \ldots, T_n \} \) be a collection of tables, and let \( Q \) be an NL question that can be independently answered by one or more tables
\( \mathcal{T}_Q = \{ T_{Q_1}, \ldots, T_{Q_m} \} \subseteq \mathcal{T} \), where \( m \geq 1 \).
The goal of a table discovery system \( S \) is to retrieve all such relevant tables:
\( \hat{\mathcal{T}}_Q = S(Q, \mathcal{T}) \approx \mathcal{T}_Q \).
The case \(m=1\) reduces to single-table discovery.
\end{definition}

\begin{example}
The question ``What is the GDP of France in 2020?'' can be answered independently by a World Bank table or an IMF table, each providing valid but slightly different statistics.  
\end{example}

\begin{definition}[Multi-Table Join Discovery]
Let \( \mathcal{T}\!=\!\{ T_1, \ldots, T_n \} \) be a collection of tables, and let \( Q \) be an NL question whose answer requires combining information across multiple tables \( \mathcal{T}_Q\!=\!\{ T_{Q_1}, \ldots, T_{Q_m} \} \allowbreak \subseteq \mathcal{T} \). 
The goal of a table discovery system \( S \) is to identify all such tables that jointly contribute to answering \( Q \): $\hat{\mathcal{T}}_Q = S(Q, \mathcal{T}) \approx \mathcal{T}_Q$.
\end{definition}

\begin{example}
The question ``Which authors have published papers in both SIGMOD and VLDB?'' requires joining an \emph{Authorship} table (author--paper relationships) with a \emph{Venue} table (paper--conference mappings).  
Neither table alone suffices to produce the complete answer.
\end{example}

\begin{definition}[Tabular Question Answering]
Let \(\mathcal T=\{T_1,\ldots,T_n\}\) be a collection of tables and
\(Q\) an NL question. Let \(A_Q\) denote the ground-truth answer derived
from the relevant tables \(\mathcal T_Q\subseteq\mathcal T\), possibly
through filtering, aggregation, or joins. A tabular question answering
system \(S\) aims to produce
\(
\hat A_Q = S(Q,\mathcal T) \approx A_Q
\).
\end{definition}

\begin{example}
The question ``What was the total sales amount for product `shoes' in 2024?'' requires retrieving the relevant Sales table, filtering the corresponding records, and aggregating their sales amounts.
\end{example}

In this work, we primarily focus on the multi-table independent and multi-table join settings introduced above. 
We additionally consider downstream question answering: after the relevant tables are identified, an NL2SQL agent extracts fine-grained answers from the discovered tables. This evaluation measures whether the retrieved tables provide sufficient evidence for downstream analytical tasks.

\section{\model}
\label{sec:octopus}

\model is a lightweight agentic framework for discovering relevant tabular data from large and heterogeneous table collections.
Its design is motivated by a tension between two common paradigms for data discovery.
Index-based retrieval systems invest substantial computation and storage in offline preprocessing to provide efficient access through a retrieval mechanism fixed in advance.
At the other extreme, general-purpose coding agents can directly inspect raw data through tools such as file search, shell commands, or executable code, but lack corpus-specific guidance and may require repeated trial-and-error exploration over a large data lake. \model is designed to occupy the middle ground between these two paradigms.

The central principle of \model is to separate semantic guidance from data inspection.
Semantic information that is reusable across queries but expensive to rediscover repeatedly is precomputed as lightweight column-level metadata.
In contrast, table contents remain directly accessible at query time and are inspected only when demanded.
Given access to the necessary information sources through a small set of high-leverage primitives, a backbone LLM is responsible for deciding how they should be used according to the query and intermediate observations, rather than imposing a fixed retrieval pipeline.

Figure~\ref{fig:octopus_overview} illustrates the overall architecture for \model.
We first describe the design principles and lightweight offline preparation in Sections~\ref{sec:octopus_overview} and~\ref{sec:offline_preparation}.
We then introduce the agentic data-access mechanisms in Section~\ref{sec:agentic_discovery}, followed by their instantiation for independent and join-based multi-table discovery in Section~\ref{sec:multi_table_discovery}.
Finally, Section~\ref{sec:cell_level_qa} describes how the discovered tables can be reused for downstream question answering via NL2SQL.

\subsection{Design Overview}
\label{sec:octopus_overview}

Let $\mathcal{T}=\{T_1,\ldots,T_n\}$ denote a collection of tables and let $Q$ be a natural-language question.
\model receives $Q$ together with access to $\mathcal{T}$ and aims to identify either a set of independently relevant tables or a group of tables that jointly support the answer, depending on the discovery setting defined in Section~\ref{section:preliminaries}.

In general, \model follows a thin-harness design.
The harness does not prescribe a fixed sequence of retrieval stages.
Instead, it provides the LLM agent with a compact collection of complementary access mechanisms:
\begin{itemize}[leftmargin=*]
    \item \textbf{Semantic schema access}, which retrieves columns and tables according to the semantic meaning of arbitrary natural-language phrases;
    \item \textbf{Direct tabular data access}, which allows the agent to inspect schemas, values, rows, and query results in the underlying tables;
    \item \textbf{General-purpose execution}, which allows open-ended exploration through shell commands when the predefined tabular primitives are insufficient.
\end{itemize}
The backbone LLM receives the original user question, the tool definitions, and observations returned from previous tool calls.
At each step, it determines which operation to invoke next, observes the execution result, and continues until sufficient evidence has been collected to produce the discovery result.
This design allows different questions to induce different execution trajectories, which are determined online by the agent.

The design therefore separates two roles that are often coupled in conventional retrieval systems.
The offline component provides lightweight semantic guidance, while the online agent remains free to inspect arbitrary raw data and adapt its retrieval strategy to the current question.
This separation is particularly important in tabular data lakes, where schemas are relatively compact but cell contents may be orders of magnitude larger and considerably more dynamic.

\subsection{Lightweight Offline Preparation}
\label{sec:offline_preparation}

The main problem of a general-purpose agent is that shell or file-access operations provide little semantic guidance.
Pure lexical search via \texttt{grep} may therefore require multiple reformulations before the agent locates the relevant schema.
\model addresses this problem by precomputing lightweight column-level representations that expose schema semantics directly to the agent.

\subsubsection*{\textbf{Column-Semantic Index}} 
Let $\mathcal{C}=\{c_1,\ldots,c_N\}$ be the set of column headers appearing in data lake $\mathcal{T}$.
\model encodes each column header once using a lightweight text encoder $f_{\mathrm{enc}}$: $\mathbf{h}_i=f_{\mathrm{enc}}(c_i)\in\mathbb{R}^{d}$,
and stores the resulting embedding matrix $\mathbf{H}=[\mathbf{h}_1;\ldots;\mathbf{h}_N] \in\mathbb{R}^{N\times d}$.

This column-semantic index is the only offline structure required for independent table discovery.
Unlike approaches that encode rows, cells, or table-level textual narrations, \model does not construct a dense content index over the data lake.
At query time, the agent may invoke semantic search with any natural-language phrase $q$ that it considers informative.
The phrase is encoded using the same encoder as $\mathbf{z}_q=f_{\mathrm{enc}}(q)$
and matched against the precomputed column representations using cosine similarity as $s(q, c_i) = (\mathbf{z}_q^\top \mathbf{h}_i) / (\Vert{}\mathbf{z}_q\Vert{}_2 \Vert{}\mathbf{h}_i\Vert{}_2)$.
The system returns the highest-scoring column names satisfying the configured similarity criterion, together with the tables containing those columns. The maximum number of unique column names returned and the similarity threshold are exposed as tool arguments and can be adjusted by the agent. The agent subsequently decides how to use the retrieved evidence.

In our implementation, we use BGE~\cite{bge} as the lightweight encoder.
The semantic index is computed once for a table collection and reused for all subsequent questions.

\subsubsection*{\textbf{Joinability Metadata}}

Identifying joinable columns in heterogeneous data lakes remains a non-trivial problem~\cite{ver, metam}, as semantically corresponding columns frequently use different names.
To support such cases without constructing a full content index, \model precomputes a lightweight MinHash~\cite{minhash} signature for each column based on its set of unique values. Specifically, for the distinct value set $V(T,c)=\{v \mid v \text{ occurs in column } c \text{ of } T\}$ of column $c$ in table $T$, a MinHash signature $\mathbf{m}(T,c)=\mathrm{MinHash}(V(T,c))$ provides an inexpensive approximation to value-set overlap and is used to identify candidate joinable columns across tables.

In this way, \model contains two forms of lightweight offline preparation for the join-based setting.
Column embeddings capture schema semantics, while MinHash signatures capture approximate column joinability.
It is worth noting that neither structure stores dense representations of individual rows or cells, and both structures can be precomputed with low computational overhead, as evaluated in detail in Section~\ref{section:evaluation}.

\subsection{Agent Harness for Data Discovery}
\label{sec:agentic_discovery}

The offline structures described above are exposed to the LLM through a complementary set of information-access primitives: semantic schema access provides broad guidance, direct tabular data access exposes precise information from the raw data, and general-purpose execution covers operations that cannot be anticipated by the harness.
Table~\ref{tab:octopus_tools} summarizes the main tool families. 

\begin{table}[t]
    \centering
    \small
    \caption{Main information-access primitives exposed by \model. Join-specific primitives are enabled for multi-table join-based discovery.}
    \label{tab:octopus_tools}
    \begin{tabular}{p{0.32\linewidth} p{0.6\linewidth}}
        \toprule
        \textbf{Primitive} & \textbf{Purpose} \\
        \midrule
        \multicolumn{1}{>{\columncolor{light-gray}}l}{\textbf{General}} & \multicolumn{1}{l}{} \\
        Semantic column search &
        Retrieve columns semantically related to an agent-specified natural-language phrase. \\

        Value search &
        Search raw table contents for a particular value. \\

        Table inspection &
        Inspect table schemas, column domains, and representative rows. \\

        SQL execution &
        Execute SQL queries over candidate tables. \\

        Shell execution &
        Perform arbitrary programmatic operations over the data lake directory. \\

        \midrule
        \multicolumn{1}{>{\columncolor{light-gray}}l}{\textbf{Join-specific}} & \multicolumn{1}{l}{} \\
        Join-column lookup &
        Find tables containing a column name. \\

        Joinability search &
        Retrieve columns in other tables with high estimated value overlap with MinHash. \\

        Overlap verification &
        Compute the value overlap between a suspected pair of join columns. \\
        \bottomrule
    \end{tabular}
\end{table}

\subsubsection*{\textbf{Semantic Schema Access}}

The column-semantic index introduced in Section~\ref{sec:offline_preparation} provides the primary source of corpus-level guidance.
The corresponding semantic-search primitive accepts a natural-language phrase supplied by the agent and returns semantically related column names, their similarity scores, and all tables containing those columns.
The agent may issue multiple semantic searches, alter a search phrase after inspecting intermediate results, or bypass semantic search entirely if lexical or structural evidence is already sufficient.
This flexibility is important when table schemas contain abbreviations, overloaded terms, or domain-specific conventions that are difficult to resolve in a single retrieval operation.

\subsubsection*{\textbf{Direct Tabular Data Access}}

Semantic similarity identifies plausible locations in the data lake, but it does not by itself establish that a candidate table satisfies the semantics of the question.
\model therefore gives the agent direct access to the underlying tabular data to search for specific value mentions, check table schemas, or execute SQL on the data lake. 
Specifically, the join-based setting requires three additional join-specific tools, which form a progressive join-exploration strategy: exact schema correspondence provides the cheapest signal; MinHash-based retrieval broadens the search to differently named but value-compatible columns; and exact value comparison verifies a small number of promising candidates. 
Combined together, these primitives allow the agent to move from broad semantic candidates to increasingly specific evidence without requiring the entire table collection to be indexed in advance.

\subsubsection*{\textbf{General-Purpose Execution}}

A fixed collection of structured tools cannot anticipate every operation that may be useful for data discovery.
Therefore, \model additionally exposes a general-purpose \texttt{bash} interface operating over the data lake directory to perform a custom analysis not directly supported by the structured primitives.
Importantly, general-purpose execution complements rather than replaces the semantic schema access and direct tabular data access tools.
Blind shell exploration over a large data lake can be expensive because the agent must repeatedly guess lexical forms and inspect broad regions of the corpus.
The semantic column index provides targeted guidance for common semantic retrieval needs, while structured tabular tools simplify frequent inspection operations.
The shell interface preserves open-endedness when neither is sufficient.
This combination is central to the thin-harness philosophy of \model: the harness exposes a compact set of high-leverage capabilities without constraining the agent to a closed retrieval procedure.

\subsection{Multi-Table Discovery}
\label{sec:multi_table_discovery}

\model is given the natural-language question $Q$, the offline column-semantic index introduced in Section~\ref{sec:offline_preparation}, and access to the primitives discussed in Section~\ref{sec:agentic_discovery}.
The same agent harness is used for both multi-table discovery settings introduced in Section~\ref{section:preliminaries}, where independent discovery seeks all tables that can satisfy the question individually, and join-based discovery seeks a collection of mutually joinable tables whose information is jointly sufficient.

\subsubsection*{\textbf{Independent Discovery}}
\label{sec:independent_discovery}

For independent discovery, the goal is to identify a set of tables in the data lake where each of them independently contains the necessary information required by $Q$.
\model does not impose a fixed target cardinality or require the same number of retrieval steps for every question. Questions with a single relevant table and questions with many independently answerable tables are handled through the same interface. The agent terminates once it determines that the accumulated evidence is sufficient and outputs the discovered table set. This setting highlights the advantage of adaptive access over a fixed top-$k$ retrieval pipeline~\cite{pneuma}, where the number and type of operations can vary according to the ambiguity of the question and the structure of the underlying data rather than being predetermined by the system.

\subsubsection*{\textbf{Join-Based Discovery}}
\label{sec:join_discovery}

For the join-based discovery, no individual table contains all information required by the question. \model must instead identify a group of tables in the data lake that collectively contain the required information and can be meaningfully joined. As discussed in Section~\ref{sec:agentic_discovery}, the agent retains all primitives available in independent discovery and is additionally given three join-specific access mechanisms.
As elsewhere in \model, this progression is not encoded as a mandatory sequence. The agent decides when each type of evidence is required based on the current question and its intermediate findings.

\subsection{Application to Downstream NL2SQL}
\label{sec:cell_level_qa}

Table discovery identifies where the information required by a question resides, whereas downstream analytical tasks may ultimately require query results rather than table identities. We therefore additionally consider an optional NL2SQL stage over the discovered tables. Given the user question together with the retrieved tables and their schemas, a separate question-answering agent generates and executes SQL to extract the final answer. This downstream stage allows us to evaluate whether differences in table discovery translate into differences in answer-level performance.

\section{Multi-Table Data Discovery Benchmark}\label{section:table_discovery_benchmark}

\begin{figure*}[!t]
\centering
\includegraphics[width=0.9\linewidth]{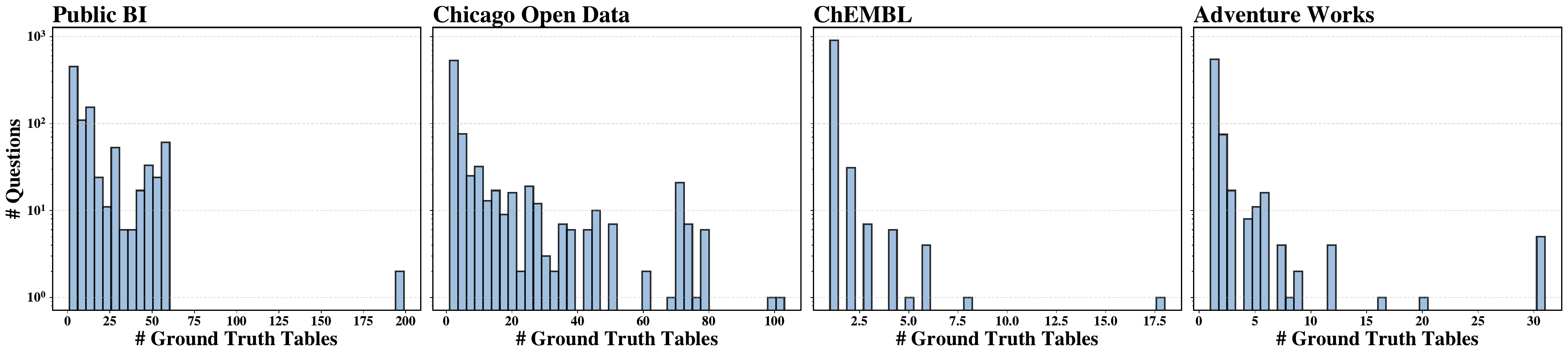}
\centering
\caption{Distribution of the number of ground-truth tables per question in the independent benchmark.}\label{figure:nojoin_dis}
\end{figure*}

\begin{figure*}[!t]
\centering
\begin{subfigure}{0.42\textwidth}
    \includegraphics[width=\linewidth]{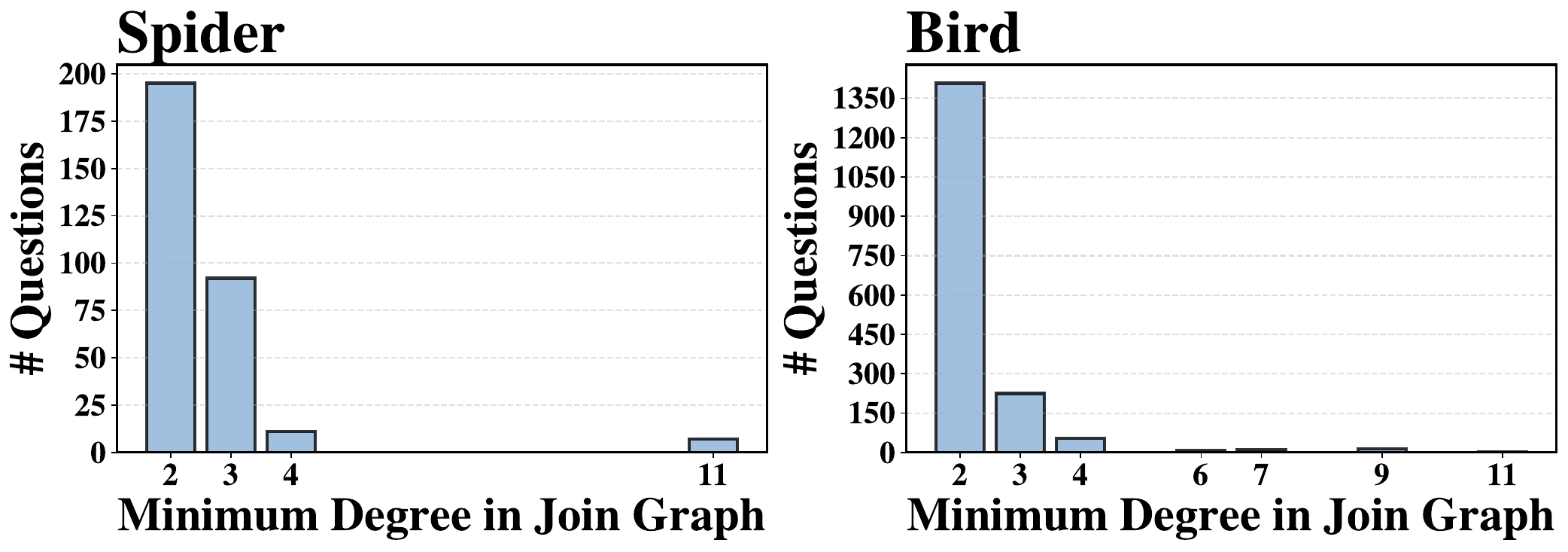}
    \subcaption{Minimum degree distribution.}
\end{subfigure}
\quad
\begin{subfigure}{0.42\textwidth}
    \includegraphics[width=\linewidth]{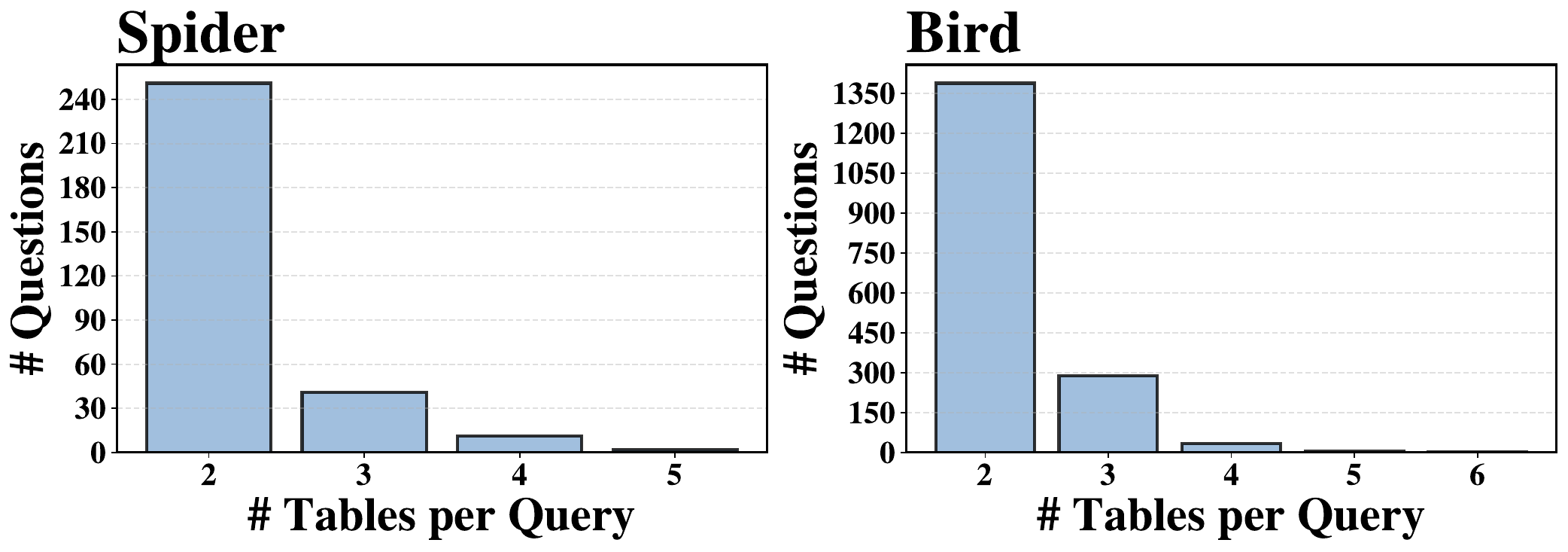}
    \subcaption{Number of answer tables distribution.}
\end{subfigure}
\caption{Dataset distribution on join-based benchmark.}
\label{figure:join_dis}
\end{figure*}

Existing tabular data discovery benchmarks suffer from several limitations.  
Traditional table question answering benchmarks~\cite{opendtr, fetaqa, qatch} scope the task to question answering over a single given table thus assuming that the relevant table is already known.
In contrast, data discovery benchmarks~\cite{cmdbench, pneuma} operate over a collection of tables, but they suffer from other issues: some incur high false-negative rates~\cite{cmdbench}, since multiple tables beyond the annotated ground truth may also correctly answer the question; others incur high false-positive rates~\cite{pneuma}, since certain questions may not be answerable by any table in the collection. 
Existing data-discovery benchmarks primarily focus on independently sufficient answer tables, whereas NL2SQL benchmarks~\cite{spider, bird} include join-based questions but assume that the relevant database is already given. Neither setting directly evaluates the discovery of a jointly required table set from a large heterogeneous table collection. Furthermore, existing data-discovery benchmarks primarily evaluate table-level retrieval and do not jointly assess the downstream answers obtainable from the discovered evidence.

To address these issues, we introduce a new benchmark for tabular data discovery with two complementary components:  
(1) independent benchmark, where one or more tables can independently provide valid answers for the questions; and  
(2) join-based benchmark, where multiple tables must be combined together to answer the questions.

\subsection{Independent Benchmark}

The most relevant benchmark to our setting is Pneuma~\cite{pneuma}, which provides four datasets containing roughly 1,000 NL questions each. In Pneuma, each record is generated from a SQL-based intermediate representation together with a corresponding NL question. Additional answer tables are annotated by checking whether they satisfy the query metadata. However, inspection of the released construction code shows that only equality predicates (``='') in the filtering clause are checked, while other comparison operators, such as ``$<$'' and ``$>$'', are ignored. This procedure can therefore introduce false-positive answer-table annotations when not all query constraints are verified.

To obtain more execution-consistent annotations, we revisit all four Pneuma datasets: Public BI, Chicago Open Data, ChEMBL, and Adventure Works. For each record, we reconstruct an executable SQL query from the released SQL metadata, including the \texttt{SELECT}, \texttt{WHERE}, \texttt{GROUP BY}, \texttt{HAVING}, \texttt{ORDER BY}, and \texttt{LIMIT} clauses when present. For quoted-string equality predicates, we use \texttt{ILIKE} instead of ``='' to reduce sensitivity to capitalization. We then execute the reconstructed query on each answer table annotated in the original benchmark. Tables on which the query cannot produce valid answer are removed from the answer-table set, and questions for which the query fails on all annotated answer tables are discarded. For downstream question answering, we concatenate the query outputs obtained from all retained answer tables. These records are subsequently normalized and deduplicated to form the reference answer for each question.

Although the resulting benchmark is designed for multi-table independent discovery, some questions have only one valid answer table after this refinement. Such cases correspond to the single-table special case of independent discovery and are therefore retained in the benchmark. Figure~\ref{figure:nojoin_dis} shows the distribution of the number of answer tables per question.

\subsection{Join-based Benchmark}

To construct a dataset where answering a question requires combining multiple tables, we need databases containing joinable tables and questions that reference columns across multiple joinable tables. Existing NL2SQL benchmarks such as Spider~\cite{spider} and Bird~\cite{bird} provide NL questions over databases with primary key–foreign key (PK-FK) relationships. However, as NL2SQL benchmarks, their scope is limited to a single database at a time involving only a small number of relevant tables. 
Moreover, in the data lake scenario we study, explicit PK–FK relationships are often not available.

We construct both join-based datasets from the training splits of Spider and Bird. First, we flatten all tables from all databases in the training split into a single global table collection. We then randomly shuffle the resulting CSV files and rename them sequentially as \texttt{table\_0.csv} and so on. This removes the original database boundaries and table identities from the agent-visible corpus, while preserving column names and cell contents. PK-FK relationships are not exposed to the discovery system; the original database structure, PK-FK annotations, and gold SQL queries are retained only for benchmark construction and ground-truth derivation.
Second, we filter out questions whose associated SQL queries do not involve explicit \texttt{JOIN} operations, and if multiple questions correspond to the same SQL, we retain only one. 
Third, we further discard queries with overly simple joins to make the datasets more challenging. Specifically, for the latent PK-FK joinable relationship, if the minimum degree of any table involved in the query is less than 2, we exclude the query. This filtering increases join-partner ambiguity by ensuring that each query-relevant table has multiple candidate neighbors in the latent join graph. 
For downstream question answering, we execute the original gold SQL query on the corresponding source database and use the resulting records as the reference answer.

Figure~\ref{figure:join_dis} reports the distributions of (1) the minimum degree of query-related tables in terms of latent PK-FK join graph and (2) the number of tables per query for the refined Spider and Bird datasets. These statistics demonstrate that the retained queries are both structurally diverse and sufficiently challenging, forming a more suitable benchmark for multi-table discovery.


\begin{table}[!t]
\centering
\caption{Statistics of datasets used in the paper.}\label{table:datasets}
\scalebox{0.78}{
\begin{tabular}{l|cccccc}
\toprule
&&Avg.&Avg. &&\\
{Name}  & {\#Tables} & {\# Rows} & {\# Attributes} & {Size} & {\# Questions} \\
\midrule
\multicolumn{1}{>{\columncolor{light-gray}}l}{\textbf{Independent}} & \multicolumn{5}{l}{} \\
Public BI & 203 & 20 & 66 & 2.2 MB & 951 \\
Chicago Open Data & 802 & 2,812 & 17 & 829 MB & 835 \\
ChEMBL & 78 & 5,161 & 7 & 67 MB & 958 \\
Adventure Works & 88 & 9,127 & 8 & 102 MB & 693 \\
\midrule
\multicolumn{1}{>{\columncolor{light-gray}}l}{\textbf{Join-based}} & \multicolumn{5}{l}{} \\
Spider & 782 & 1,879 & 5 & 95 MB & 305 \\
Bird & 427 & 830,504 & 7 & 15 GB & 1,716 \\
\bottomrule
\end{tabular}
}
\end{table}

\section{Evaluation}\label{section:evaluation}
In this section, we evaluate the effectiveness of \model by answering the following research questions:

\begin{itemize}
\item[\textbf{RQ1:}] How effective is \model at table-level discovery in both independent and join-based settings?
\item[\textbf{RQ2:}] Do improvements in table discovery translate into better downstream NL2SQL performance?
\item[\textbf{RQ3:}] How efficient is \model in terms of offline preprocessing time and storage, as well as online latency and API cost?
\item[\textbf{RQ4:}] Can the Column-Semantic Index serve as a reusable retrieval primitive for off-the-shelf coding agents?
\item[\textbf{RQ5:}] How does the choice of backbone LLM affect the effectiveness and efficiency of \model?
\item[\textbf{RQ6:}] How does \model use its different information-access mechanisms during table discovery?
\end{itemize}

\subsection{Experimental Setup}
\subsubsection*{\textbf{Datasets}}
We conduct experiments on the benchmarks introduced in Section~\ref{section:table_discovery_benchmark}. Specifically, for independent benchmark, we have business intelligence dataset Public BI~\cite{public}, government open dataset Chicago Open Data~\cite{chicago}, biochemical dataset ChEMBL~\cite{chembl}, and enterprise operation dataset Adventure Works~\cite{adventure}. For join-based benchmark, we have two large-scale complex and cross-domain semantic parsing and NL2SQL datasets Spider~\cite{spider} and Bird~\cite{bird}. Table~\ref{table:datasets} shows the statistics for these datasets.

\subsubsection*{\textbf{Baselines}}
We compare \model against tabular data discovery methods and general coding agents that generalize to arbitrary table collections without task-specific training for table retrieval:
\begin{itemize}
\setlength{\leftskip}{-2em}
\item \textbf{Full-Text Search}: A naive approach that treats table contents as plain text. BM25~\cite{bm25} is leveraged to retrieve tables based on string-level match between table contents and NL user questions.
\item \textbf{LlamaIndex}~\cite{llamaindex}: An approach that treats each row as a record and encodes records into dense vectors with a sentence encoder. Tables are retrieved by aggregating record-level similarities to the query embedding.
\item \textbf{Pneuma}~\cite{pneuma}: A state-of-the-art retrieval-augmented generation (RAG)-based approach that combines full-text search and vector search with LLM-based relevance judgment to retrieve the most relevant tables.
\item \textbf{Claude Code}~\cite{claudecode}: A state-of-the-art coding agent that performs agentic search through direct corpus interaction.
\end{itemize}

For downstream question answering, we compare \model with Claude Code and apply the same NL2SQL agent with Claude Haiku 4.5~\cite{haiku4.5} as the backbone to the tables returned by each discovery method to ensure a fair comparison. 

\subsubsection*{\textbf{Evaluation Metrics}}


\begin{table*}[t]
\centering
\caption{Table retrieval results on the independent benchmark in terms of macro Precision, Recall, and F1. T/O indicates that the method did not complete within 20 hours and is therefore omitted.}
\label{tab:table-level-independent}
\scalebox{1.00}{
\begin{tabular}{l|ccc|ccc|ccc|ccc}
\toprule
& \multicolumn{3}{c|}{\textbf{Public BI}}
& \multicolumn{3}{c|}{\textbf{Chicago Open Data}}
& \multicolumn{3}{c|}{\textbf{ChEMBL}}
& \multicolumn{3}{c}{\textbf{Adventure Works}} \\
\cmidrule(lr){2-4}
\cmidrule(lr){5-7}
\cmidrule(lr){8-10}
\cmidrule(lr){11-13}

\textbf{Method}
& \textbf{P} & \textbf{R} & \textbf{F1}
& \textbf{P} & \textbf{R} & \textbf{F1}
& \textbf{P} & \textbf{R} & \textbf{F1}
& \textbf{P} & \textbf{R} & \textbf{F1} \\
\midrule

\multicolumn{1}{>{\columncolor{light-gray}}l|}{\textbf{Full-Text Search}}
& \multicolumn{12}{l}{} \\

$k=1$
& 0.493 & 0.177 & 0.222
& 0.044 & 0.040 & 0.041
& 0.146 & 0.142 & 0.143
& 0.286 & 0.247 & 0.257 \\

$k=5$
& 0.435 & 0.452 & 0.353
& 0.019 & 0.066 & 0.027
& 0.096 & 0.459 & 0.157
& 0.126 & 0.491 & 0.189 \\

$k=10$
& 0.379 & 0.596 & 0.363
& 0.013 & 0.089 & 0.021
& 0.080 & 0.742 & 0.142
& 0.089 & 0.639 & 0.145 \\

\midrule

\multicolumn{1}{>{\columncolor{light-gray}}l|}{\textbf{LlamaIndex}}
& \multicolumn{12}{l}{} \\

$k=1$
& 0.423 & 0.131 & 0.163
& T/O & T/O & T/O
& T/O & T/O & T/O
& T/O & T/O & T/O \\

$k=5$
& 0.417 & 0.204 & 0.214
& T/O & T/O & T/O
& T/O & T/O & T/O
& T/O & T/O & T/O \\

$k=10$
& 0.412 & 0.239 & 0.234
& T/O & T/O & T/O
& T/O & T/O & T/O
& T/O & T/O & T/O \\

\midrule

\multicolumn{1}{>{\columncolor{light-gray}}l|}{\textbf{Pneuma}}
& \multicolumn{12}{l}{} \\

$k=1$
& 0.593 & 0.155 & 0.199
& \textbf{0.434} & 0.287 & 0.304
& 0.770 & 0.746 & 0.752
& 0.641 & 0.558 & 0.576 \\

$k=5$
& 0.601 & 0.414 & 0.398
& 0.308 & 0.493 & 0.299
& 0.296 & 0.950 & 0.439
& 0.330 & 0.896 & 0.446 \\

$k=10$
& 0.579 & 0.564 & 0.470
& 0.243 & \textbf{0.566} & 0.255
& 0.155 & \textbf{0.970} & 0.262
& 0.192 & \textbf{0.948} & 0.293 \\

\midrule

Claude Code
& 0.476 & 0.494 & 0.473
& 0.140 & 0.170 & 0.143
& 0.390 & 0.455 & 0.403
& 0.417 & 0.483 & 0.432 \\

\model
& \textbf{0.916} & \textbf{0.945} & \textbf{0.913}
& 0.392 & 0.433 & \textbf{0.389}
& \textbf{0.915} & 0.954 & \textbf{0.921}
& \textbf{0.882} & 0.922 & \textbf{0.889} \\

\bottomrule
\end{tabular}
}
\end{table*}

Unlike prior table discovery evaluations that only use hit rate~\cite{pneuma}, our settings require metrics that reflect multi-table answers and answer-level evidence.
For table discovery, we compute Precision, Recall, and F1 for each question based on the predicted and ground-truth table sets, and report
their macro averages:
\(
P=\frac{1}{N}\sum_i P_i,\;
R=\frac{1}{N}\sum_i R_i,\;
F1=\frac{1}{N}\sum_i F1_i.
\)
For the join-based benchmark, we additionally report Success Rate, defined as the fraction of questions for which all ground-truth tables are retrieved, and Exact Match, defined as the fraction for which the predicted and ground-truth table sets are identical.
For downstream question answering, we compare the answer records produced by the NL2SQL agent against the reference answer associated with each question. Both reference and predicted records are normalized and deduplicated before comparison. We then compute record-level Precision, Recall, and F1 for each question and report their macro averages.

\subsubsection*{\textbf{Implementation Configurations}}
For agent-based methods Claude Code and \model, we set the execution cap for each question to be 300s and 30 turns, and use Claude Haiku 4.5 as the default backbone LLM with temperature 0.0. We also use lightweight BGE model~\cite{bge} as the text encoder for header embeddings for \model.
For non-agentic retrieval systems Full-Text Search, LlamaIndex, and Pneuma, the number of returned tables \(k\) must be specified in advance. We therefore evaluate \(k\in\{1,5,10\}\) on the independent benchmark and \(k\in\{1,2,5,10\}\) on the join-based benchmark.
For Pneuma in particular, we set the balancing parameter \(\alpha=0.5\) and the initial-retrieval multiplier \(n=5\), and use Qwen2.5-7B-Instruct~\cite{qwen2.5} as the LLM judge, following the settings recommended in the original paper~\cite{pneuma}.
All experiments are conducted on a Linux server with one A6000 GPU and Python 3.12.2.

\subsection{Table Discovery Evaluation}
In this section, we evaluate \model for table discovery on both independent and join-based benchmarks.

\subsubsection*{\textbf{Independent Benchmark}}

We report macro-Precision, Recall, and F1 on the four independent-discovery datasets in Table~\ref{tab:table-level-independent}.
For non-agentic retrieval methods that require the number of returned tables $k$ to be specified in advance, we evaluate $k\in\{1,5,10\}$.
In contrast, Claude Code and \model determine the number of returned tables adaptively during the agentic search process.
Since LlamaIndex requires encoding every row in the corpus, it does not complete within 20 hours on Chicago Open Data, ChEMBL, and Adventure Works; we therefore report these cases as ``T/O''.

Overall, \model achieves the highest macro-F1 on all four datasets.
The gains are particularly pronounced on Public BI and Adventure Works, where \model improves macro-F1 over the strongest baseline by 0.440 and 0.313, respectively.
Moreover, \model substantially outperforms the general-purpose Claude Code agent on every dataset, indicating that direct agentic access to the raw corpus alone is less effective for reliable table discovery.
The lightweight semantic guidance and structured tabular access provided by \model allow the agent to narrow the search space while retaining the flexibility to inspect and verify candidate tables adaptively.

The results also highlight an inherent limitation of fixed-$k$ retrieval.
Increasing $k$ can improve recall by returning more candidate tables, but often at the expense of substantially lower precision.
For example, on ChEMBL, Pneuma with \(k=10\) achieves slightly higher recall than \model (0.970 vs. 0.954). However, its precision drops to 0.155, compared with 0.915 for \model, resulting in a much lower F1 score (0.262 vs. 0.921). Thus, aggressively increasing the retrieval set can improve recall without yielding accurate table discovery.
More fundamentally, requiring users to specify $k$ is poorly suited to data discovery as the number of relevant tables varies substantially across questions and is generally unknown beforehand for an end user.
By contrast, agent-based methods Claude Code and \model do not assume a predetermined answer cardinality to continue gathering and verifying evidence until the agent determines that the relevant table set is sufficiently supported.
This adaptive stopping behavior allows \model to maintain a substantially better precision–recall balance across datasets with heterogeneous answer cardinalities.


\begin{table*}[t]
\centering
\caption{Table retrieval results on the join-based benchmark in terms of macro Precision, Recall, F1, Success Rate, and Exact Match.}
\label{tab:table-level-join}
\scalebox{1.00}{
\begin{tabular}{l|ccccc|ccccc}
\toprule
& \multicolumn{5}{c|}{\textbf{Spider}}
& \multicolumn{5}{c}{\textbf{Bird}} \\
\cmidrule(lr){2-6}
\cmidrule(lr){7-11}

\textbf{Method}
& \textbf{P} & \textbf{R} & \textbf{F1} & \textbf{Succ} & \textbf{EM}
& \textbf{P} & \textbf{R} & \textbf{F1} & \textbf{Succ} & \textbf{EM} \\
\midrule

\multicolumn{1}{>{\columncolor{light-gray}}l|}{\textbf{Full-Text Search}}
& \multicolumn{10}{l}{} \\

$k=1$
& 0.036 & 0.018 & 0.024 & 0.000 & 0.000
& 0.027 & 0.013 & 0.018 & 0.000 & 0.000 \\

$k=2$
& 0.030 & 0.028 & 0.029 & 0.007 & 0.007
& 0.025 & 0.024 & 0.024 & 0.000 & 0.000 \\

$k=5$
& 0.054 & 0.125 & 0.075 & 0.020 & 0.000
& 0.018 & 0.040 & 0.024 & 0.004 & 0.000 \\

$k=10$
& 0.043 & 0.197 & 0.070 & 0.039 & 0.000
& 0.013 & 0.060 & 0.022 & 0.013 & 0.000 \\

\midrule

\multicolumn{1}{>{\columncolor{light-gray}}l|}{\textbf{Pneuma}}
& \multicolumn{10}{l}{} \\

$k=1$
& 0.541 & 0.259 & 0.349 & 0.000 & 0.000
& \textbf{0.480} & 0.225 & 0.305 & 0.000 & 0.000 \\

$k=2$
& 0.512 & 0.381 & 0.425 & 0.112 & 0.112
& 0.475 & 0.370 & 0.404 & 0.074 & 0.074 \\

$k=5$
& 0.324 & 0.618 & 0.421 & 0.380 & 0.000
& 0.309 & 0.566 & 0.394 & 0.290 & 0.004 \\

$k=10$
& 0.200 & \textbf{0.726} & 0.311 & 0.544 & 0.000
& 0.199 & \textbf{0.689} & 0.305 & \textbf{0.453} & 0.000 \\

\midrule

Claude Code
& 0.201 & 0.191 & 0.194 & 0.157 & 0.154
& 0.202 & 0.193 & 0.195 & 0.167 & 0.160 \\

\model
& \textbf{0.688} & 0.674 & \textbf{0.675} & \textbf{0.554} & \textbf{0.515}
& 0.455 & 0.443 & \textbf{0.445} & 0.401 & \textbf{0.385} \\

\bottomrule
\end{tabular}
}
\end{table*}


\begin{table*}[t]
\centering
\caption{Downstream question answering evaluation on the independent benchmark in terms of macro Precision, Recall, and F1.}
\label{tab:cell-level-independent}
\scalebox{1.00}{
\begin{tabular}{l|ccc|ccc|ccc|ccc}
\toprule
& \multicolumn{3}{c|}{\textbf{Public BI}}
& \multicolumn{3}{c|}{\textbf{Chicago Open Data}}
& \multicolumn{3}{c|}{\textbf{ChEMBL}}
& \multicolumn{3}{c}{\textbf{Adventure Works}} \\
\cmidrule(lr){2-4}
\cmidrule(lr){5-7}
\cmidrule(lr){8-10}
\cmidrule(lr){11-13}

\textbf{Method}
& \textbf{P} & \textbf{R} & \textbf{F1}
& \textbf{P} & \textbf{R} & \textbf{F1}
& \textbf{P} & \textbf{R} & \textbf{F1}
& \textbf{P} & \textbf{R} & \textbf{F1} \\
\midrule


Claude Code
& \textbf{0.914} & 0.411 & 0.415
& \textbf{0.910} & 0.158 & 0.158
& \textbf{0.893} & 0.313 & 0.315
& \textbf{0.872} & 0.374 & 0.360 \\

\model
& 0.813 & \textbf{0.616} & \textbf{0.625}
& 0.792 & \textbf{0.267} & \textbf{0.269}
& 0.799 & \textbf{0.628} & \textbf{0.631}
& 0.755 & \textbf{0.634} & \textbf{0.593} \\

\bottomrule
\end{tabular}
}
\end{table*}


\begin{table}[t]
\centering
\caption{Downstream question answering evaluation on the join-based benchmark in terms of macro Precision, Recall, and F1.}
\label{tab:cell-level-join}
\scalebox{1.00}{
\begin{tabular}{l|ccc|ccc}
\toprule
& \multicolumn{3}{c|}{\textbf{Spider}}
& \multicolumn{3}{c}{\textbf{Bird}} \\
\cmidrule(lr){2-4}
\cmidrule(lr){5-7}

\textbf{Method}
& \textbf{P} & \textbf{R} & \textbf{F1}
& \textbf{P} & \textbf{R} & \textbf{F1} \\
\midrule

Claude Code
& \textbf{0.870} & 0.137 & 0.134
& \textbf{0.862} & 0.085 & 0.083 \\

\model
& 0.590 & \textbf{0.482} & \textbf{0.466}
& 0.687 & \textbf{0.197} & \textbf{0.194} \\

\bottomrule
\end{tabular}
}
\end{table}


\subsubsection*{\textbf{Join-based Benchmark}}

We report the results on the two join-based datasets in Table~\ref{tab:table-level-join}.
Unlike independent discovery, answering a query in this setting requires retrieving a group of tables that collectively contains all necessary information.
Agentic methods, including Claude Code and \model, can determine the table group adaptively during the discovery process.
For non-agentic methods, the number of returned tables must instead be specified in advance; we therefore evaluate $k\in\{1,2,5,10\}$.
Pneuma is originally designed to judge whether an individual table is sufficient to answer a question.
To adapt it to the join setting, we modify its LLM judge from ``Is the table relevant to answer the question?'' to ``Are some columns in the table relevant to answer the question?''
LlamaIndex does not finish within 20 hours on either data lake and is therefore omitted.

Overall, \model achieves the highest F1 and Exact Match on both datasets,
which indicates that \model is substantially more effective at identifying a compact set of tables for multi-table reasoning, rather than merely retrieving individually plausible candidates.
For the non-agentic baselines, increasing \(k\) retrieves more ground-truth tables and therefore improves recall and Success Rate, but at the cost of introducing more irrelevant tables.

A closer look at the baselines reveals why they struggle with this trade-off.
Full-Text Search performs poorly across all metrics, with F1 remaining below $0.08$ on both datasets even at larger values of $k$.
This is expected because lexical relevance alone provides little evidence about whether independently retrieved tables can be meaningfully combined.
Pneuma performs substantially better at identifying semantically relevant tables, but its table-wise retrieval formulation still does not explicitly model whether the retrieved candidates together form the required join group.
In contrast, \model augments semantic discovery with join-specific primitives for identifying and verifying joinable columns, allowing the agent to reason over both table relevance and cross-table connectivity.
The large gap between \model and Claude Code further suggests that general-purpose agentic exploration alone is less effective in this setting, and that the additional corpus-specific and join-oriented access mechanisms provided by \model are beneficial for multi-table discovery.


\begin{figure*}[!t]
\centering
\includegraphics[width=\linewidth]{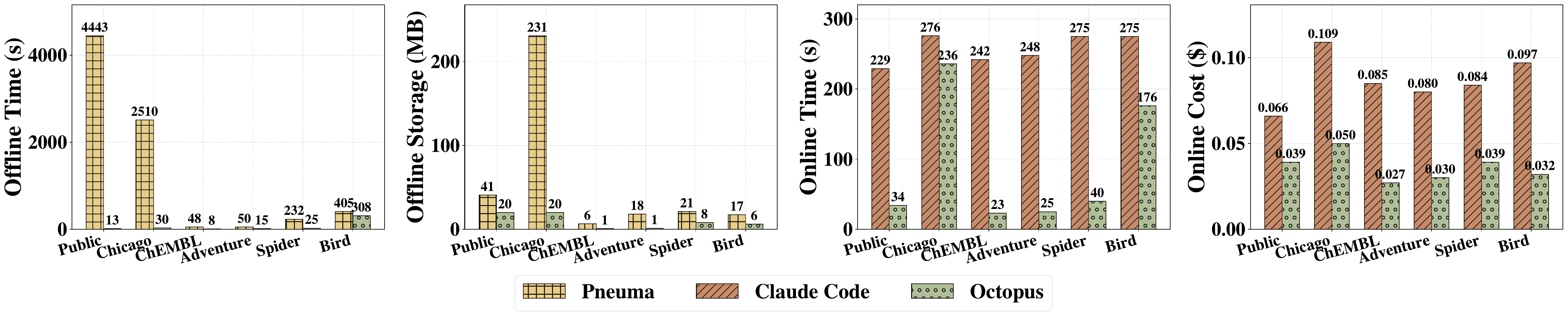}
\centering
\caption{Efficiency comparison of \model and the baselines. (Left two): Offline efficiency measured by preparation time and storage footprint. (Right two): Online efficiency measured by processing time and cost per question.}\label{figure:efficiency_all}
\end{figure*}


\subsection{NL2SQL Evaluation}

We further evaluate whether the improvement in table discovery translates into better downstream question answering by applying the same NL2SQL agent to the tables retrieved in the previous stage. For this evaluation, we compare \model with Claude Code, as both methods adaptively determine the returned table set rather than requiring a predefined retrieval cardinality \(k\).

\subsubsection*{\textbf{NL2SQL Agent}}

The NL2SQL agent follows the same thin-harness principle as \model.
Given a natural-language question and the tables retrieved in the previous stage, the agent is provided with the schemas of all candidate tables and iteratively interacts with them through lightweight tools.
Specifically, it can inspect sample rows and column values, generate and execute SQL queries over the retrieved tables using DuckDB, and refine its queries based on the execution results.
Once it identifies an answer, the agent returns the resulting query records for evaluation.
We fix Claude Haiku 4.5 as the backbone for the NL2SQL agent for both discovery methods, such that the downstream reasoning capability remains identical and the primary difference lies in the evidence supplied by the preceding table-discovery stage.

For both Claude Code and \model, we directly pass the complete table set returned by the discovery agent without imposing an additional cardinality constraint. The NL2SQL agent is restricted to these retrieved tables and cannot access other tables in the data lake. Therefore, relevant evidence missed during table discovery cannot be recovered in the downstream stage.

\subsubsection*{\textbf{NL2SQL Evaluation Results}}

Table~\ref{tab:cell-level-independent} and Table~\ref{tab:cell-level-join} report macro-Precision, Recall, and F1 for the independent and join-based benchmarks, respectively.
Overall, \model achieves higher answer-level F1 than Claude Code across all evaluated datasets, showing that its gains in table discovery translate directly into more effective downstream question answering.
The precision-recall pattern is particularly consequential in the join setting, since missing even one table participating in the required join may prevent the downstream agent from reconstructing the complete answer.
The higher recall obtained with \model indicates that its discovery stage provides the downstream agent with more complete evidence.

Notably, Claude Code achieves higher answer-level precision than \model across all datasets, but its recall is substantially lower. This suggests that retrieving a small set of highly plausible tables is insufficient when relevant evidence is omitted. By recovering a more complete set of relevant tables, \model provides the downstream NL2SQL agent with better evidence coverage, resulting in substantially higher F1 across all datasets. Since the same NL2SQL agent is used for both discovery methods and is restricted to the tables supplied by the discovery stage, these results further show that downstream reasoning cannot fully compensate for missing upstream evidence.

\subsection{Efficiency Analysis}

We compare the efficiency of \model in both the offline preparation and online processing stages.
Figure~\ref{figure:efficiency_all} reports offline preprocessing time and storage footprint against Pneuma, and per-question online execution time and API cost against Claude Code.

\subsubsection*{\textbf{Offline Efficiency}}

\model requires substantially less offline preprocessing than Pneuma across all datasets.
In particular, on Public BI, the offline preparation time decreases from $4443$ seconds for Pneuma to only $13$ seconds for \model, corresponding to a $342\times$ speedup.
In terms of storage, \model consistently maintains a smaller footprint, with the largest reduction occurring on Adventure Works, where it requires approximately $18\times$ less storage.

This efficiency follows directly from the lightweight offline design of \model.
For independent discovery, \model only embeds column headers once and reuses these representations across all subsequent queries.
Even in the join-based setting, it supplements these embeddings only with lightweight MinHash signatures for estimating column joinability.
It therefore avoids constructing dense representations over individual rows or cells.
Pneuma, by contrast, invokes an LLM to generate table-level schema narrations and constructs additional retrieval indexes over table content.
Consequently, its preprocessing cost grows with the amount of table content that must be represented and indexed, whereas the offline structures of \model remain comparatively compact.

\subsubsection*{\textbf{Online Efficiency}}

\begin{table*}[t]
\centering
\caption{Ablation study of the Column-Semantic Index tool on the independent and join-based benchmarks. For the independent setting, we report table-level retrieval performance on Public BI; for the join-based setting, we report table-level retrieval performance on Spider.}
\label{tab:embedding-ablation}
\begin{tabular}{lccc cc}
\toprule
\textbf{Variant}
& \textbf{P} & \textbf{R} & \textbf{F1}
& \textbf{Time (s)} & \textbf{Cost (\$)} \\
\midrule

\multicolumn{1}{>{\columncolor{light-gray}}l}{\textbf{Independent (Public BI)}}
& \multicolumn{5}{l}{} \\

\model
& 0.916 & 0.945 & 0.913
& 34.1 & 0.039 \\

\quad w/o Column-Semantic Index
& 0.879 ($-$4.0\%)
& 0.870 ($-$7.9\%)
& 0.855 ($-$6.4\%)
& 79.1 (+132.0\%)
& 0.050 (+28.2\%) \\

\midrule

\multicolumn{1}{>{\columncolor{light-gray}}l}{\textbf{Join-based (Spider)}}
& \multicolumn{5}{l}{} \\

\model
& 0.688 & 0.674 & 0.675
& 39.2 & 0.039 \\

\quad w/o Column-Semantic Index
& 0.625 ($-$9.2\%)
& 0.603 ($-$10.5\%)
& 0.608 ($-$9.9\%)
& 136.1 (+247.2\%)
& 0.043 (+10.3\%) \\

\bottomrule
\end{tabular}
\end{table*}

\begin{table}[t]
\centering
\caption{Effect of the Column-Semantic Index on Claude Code on the Public BI benchmark.}
\label{tab:mcp-ablation}
\scalebox{0.85}{
\begin{tabular}{lccc cc}
\toprule
\textbf{Method}
& \textbf{P} & \textbf{R} & \textbf{F1}
& \textbf{Time (s)} & \textbf{Cost (\$)} \\
\midrule

Claude Code
& 0.476 & 0.494 & 0.473
& 228.5 & 0.066 \\

\quad with Semantic Index
& 0.542
& 0.573
& 0.544
& 222.7
& 0.064 \\

\quad $\Delta$
& +13.9\%
& +16.0\%
& +15.0\%
& $-$2.5\%
& $-$3.0\% \\

\bottomrule
\end{tabular}
}
\end{table}


The lightweight semantic guidance also substantially reduces the cost of online agentic exploration.
As shown in Figure~\ref{figure:efficiency_all}, \model is faster than Claude Code on all six datasets, while simultaneously incurring lower API cost.
The largest latency reduction occurs on ChEMBL, where the average processing time decreases from $242$ seconds per question for Claude Code to $23$ seconds for \model, corresponding to a $10.5\times$ speedup.

A similar trend holds for API usage.
Across all datasets, \model reduces the per-question API cost compared with Claude Code.
These savings indicate that the efficiency gain is not merely due to faster execution of the same agent trajectory.
Rather, \model generally incurs less API cost to locate the relevant evidence.

We attribute these improvements to the role of the Column-Semantic Index as lightweight corpus-specific guidance.
A general-purpose agent such as Claude Code can inspect arbitrary files and execute flexible search operations, but without semantic knowledge of the data lake it may need to repeatedly formulate lexical queries, inspect broad regions of the corpus, and revise unsuccessful searches.
In contrast, \model can directly query the precomputed semantic index to identify columns and tables related to an agent-specified concept, and then use direct table inspection or general-purpose execution only when further verification is necessary.
The offline semantic index therefore reduces expensive blind exploration without restricting the flexibility of the agentic search process.

\subsection{Ablation Study}\label{section:ablation}

We study how the design choices of \model, particularly the Column-Semantic Index introduced in Section~\ref{sec:offline_preparation}, affect its retrieval effectiveness and efficiency.
We further investigate whether the semantic index can benefit a general-purpose coding agent, and how different backbone LLMs affect the behavior of \model.

\subsubsection*{\textbf{Column-Semantic Index}}

We first ablate the Column-Semantic Index while keeping the remaining agent harness and tools unchanged.
The results are shown in Table~\ref{tab:embedding-ablation}.
Removing the semantic index consistently degrades retrieval quality while substantially increasing online search cost.

These results show that the Column-Semantic Index serves two complementary roles.
First, it improves retrieval effectiveness by providing the agent with semantic schema evidence that may be difficult to recover through lexical search alone.
Second, it substantially reduces the amount of blind exploration required over the raw data lake.
Without the index, the agent must rely more heavily on repeated value searches, file inspection, and general-purpose execution to locate relevant schema evidence.
The larger degradation on Spider further suggests that semantic guidance becomes particularly useful in the join setting, where the agent must locate multiple complementary tables before reasoning about their joinability.

Furthermore, the Column-Semantic Index is not specific to the \model harness.
We expose the same semantic-search primitive to Claude Code through the Model Context Protocol (MCP) and evaluate it on Public BI.
Claude Code can invoke the tool during its native exploration process to retrieve query-relevant columns and their associated tables, while retaining its existing file-access and reasoning capabilities.
This integration requires no modification to the internal agent implementation, making the index a lightweight and reusable retrieval primitive for general-purpose agents.

Table~\ref{tab:mcp-ablation} shows that adding the Column-Semantic Index improves Claude Code across all retrieval metrics.
Macro-Precision increases from $0.476$ to $0.542$, Recall from $0.494$ to $0.573$, and F1 from $0.473$ to $0.544$, corresponding to a $15.0\%$ relative improvement in F1.
Importantly, this improvement does not come at the expense of efficiency: average execution time decreases slightly from $228.5$ to $222.7$ seconds, while API cost decreases from $\$0.066$ to $\$0.064$.
Thus, even for an independently developed coding agent, lightweight semantic guidance enables more accurate discovery while slightly reducing the amount of exploration required.
This result shows that the Column-Semantic Index can also benefit a general-purpose coding agent outside the \model harness, rather than serving only as an \model-specific optimization.

\subsubsection*{\textbf{Choice of LLM}}


\begin{table}[t]
\centering
\caption{Model-backbone ablation for \model on the Public BI dataset.}
\label{tab:model-ablation}
\begin{tabular}{lccc cc}
\toprule
\textbf{Model}
& \textbf{P} & \textbf{R} & \textbf{F1}
& \textbf{Time (s)} & \textbf{Cost (\$)} \\
\midrule

Claude Haiku 4.5
& 0.916 & 0.945 & 0.913
& 34.1 & 0.039 \\

Claude Sonnet 4.5
& 0.920 & 0.911 & 0.898
& 62.6 & 0.146 \\

DeepSeek V3.2
& 0.329 & 0.339 & 0.330
& 147.2 & 0.166 \\

GLM-5
& 0.728 & 0.711 & 0.707
& 79.6 & 0.182 \\

\bottomrule
\end{tabular}
\end{table}


We further evaluate the sensitivity of \model to its backbone LLM by replacing Claude Haiku 4.5 with Claude Sonnet 4.5~\cite{anthropic2025claudesonnet45}, DeepSeek-V3.2~\cite{deepseekai2025deepseekv32}, and GLM-5~\cite{glm5team2026glm5} on Public BI.
Table~\ref{tab:model-ablation} shows that a more expensive backbone does not necessarily yield better retrieval performance.

Our execution traces further indicate that these differences are strongly affected by the models' agentic control behavior, particularly their termination decisions.
The Anthropic backbones submit a final table set for more than $96\%$ of questions, whereas DeepSeek-V3.2 submits on only approximately $34\%$.
For many remaining cases, the latter continues searching or verifying candidates until the turn budget is exhausted without producing a final answer.
Therefore, the backbone comparison reflects not only semantic reasoning ability, but also how reliably a model decides that sufficient evidence has been collected and terminates the agentic search process.
Investigating how model training affects such stopping behavior is beyond the scope of this work.

\subsection{Agent Behavior Analysis}

\begin{figure}[!t]
    \centering
    \includegraphics[width=\linewidth]{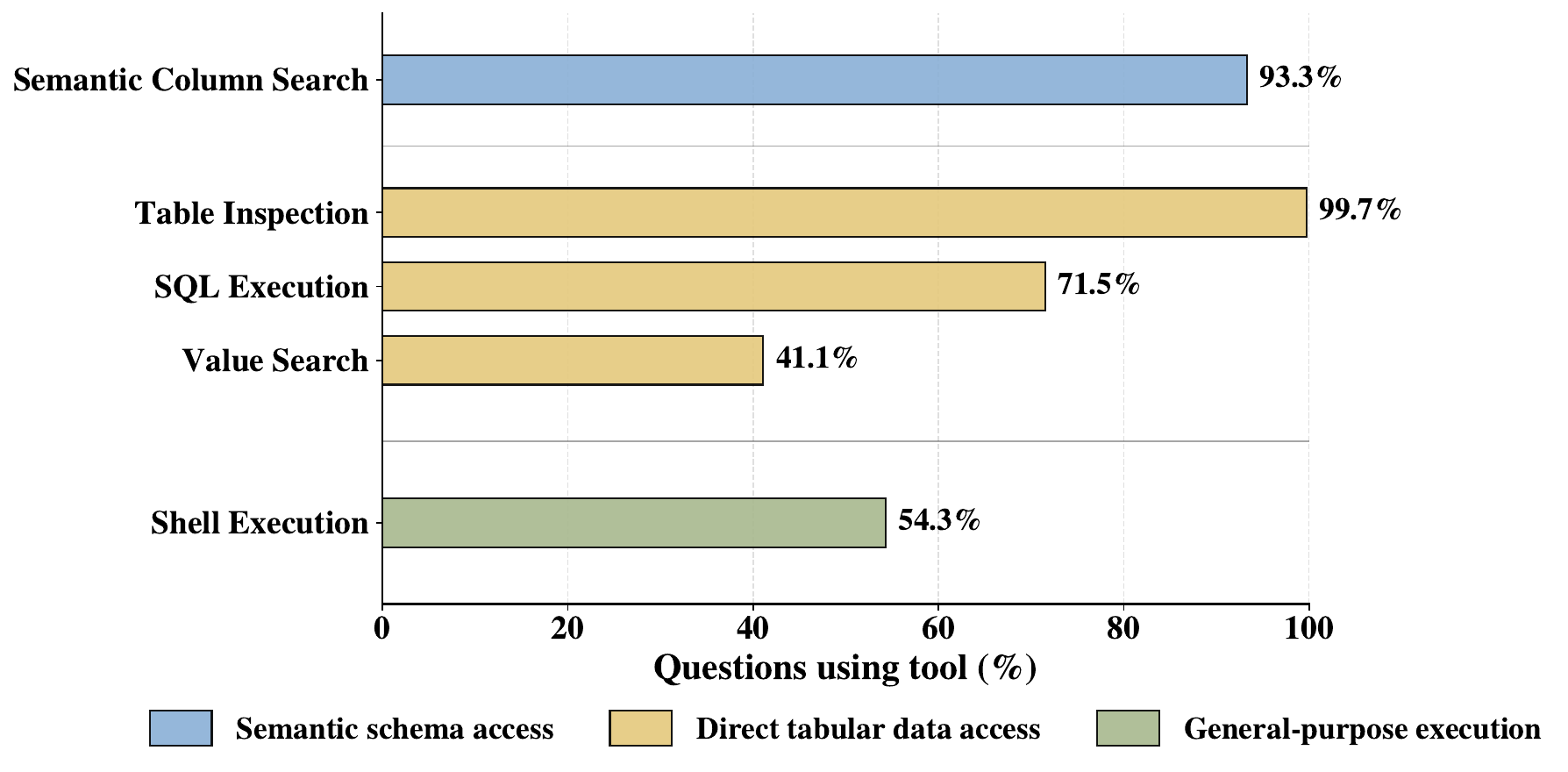}
    \caption{Tool usage frequency of \model on the Public BI benchmark.}
    \label{fig:tool-usage}
\end{figure}

To understand how \model uses the complementary access mechanisms provided by its thin harness, we analyze the complete tool-interaction traces of all 951 Public BI questions using Claude Haiku 4.5 as the backbone LLM. Figure~\ref{fig:tool-usage} summarizes how often the different information-access tools are used across questions. Direct tabular access is broadly used in practice: table inspection appears in $99.7\%$ of questions, and SQL execution appears in $71.5\%$. Semantic schema access is also widely exercised, with semantic column search used in $93.3\%$ of questions, showing that lightweight semantic guidance plays an important role in orienting the search process. General-purpose execution remains complementary but still substantial, with shell execution used in $54.3\%$ of questions. Overall, these results show that all three access mechanisms are actively used in practice while serving distinct roles: semantic search provides corpus-level guidance, direct tabular operations support candidate inspection and verification, and general-purpose execution preserves open-endedness beyond the predefined structured primitives.

\section{Related Work}\label{section:related_work}

\subsubsection*{\textbf{Tabular Data Discovery}}
Tabular data discovery aims to retrieve tables relevant to a user need from large table collections~\cite{auctus,aurum,zhang2020finding,opendtr,gtr,pneuma,gent,ver,metam}. Natural language has emerged as an increasingly common interface for broad audiences~\cite{opendtr,gtr,solo,pneuma}. However, existing systems typically incur substantial offline or online costs: many require training or schema/content indexing at scale (long offline time and large storage footprint), while others rely on expensive LLM judgments during retrieval (high online latency). Moreover, most discovery methods ultimately rank tables independently and are not designed to recover a jointly required table set, whether the relevant tables provide independently sufficient evidence or must be composed through joins.

\subsubsection*{\textbf{Tabular Question Answering}}
Tabular question answering focuses on producing answers to NL questions given one or more provided tables. Recent state-of-the-art approaches increasingly leverage LLMs for language understanding and table reasoning, including NL2SQL-style pipelines and tool-augmented prompting~\cite{nl2sql,reactable,autoprep}. While these methods target final answers, they presuppose access to the relevant tables and therefore do not address the upstream problem of discovering evidence within large, heterogeneous repositories.

\subsubsection*{\textbf{Agentic Search}}

Recent advances in LLM agents have shifted retrieval from a single-shot retrieve-then-reason paradigm toward iterative search, where an agent can interleave reasoning, tool invocation, and observations to progressively refine its information needs~\cite{react}. Recent retrieval agents further train or prompt LLMs to autonomously formulate and reformulate queries over multiple search rounds~\cite{searchr1}. While these approaches make the retrieval process adaptive, corpus access is typically still mediated by a predefined search engine or retriever. More recently, direct corpus interaction has emerged as an alternative interface for agentic search. Recent work~\cite{dci} shows that capable agents can bypass conventional retrievers and directly search raw corpora using general-purpose primitives such as grep, file reads, and shell commands. Related work similarly demonstrates that agentic keyword search can provide competitive document retrieval and question-answering performance without maintaining a vector database~\cite{subramanian2026keyword}. These results highlight the flexibility of exposing fine-grained corpus access directly to LLM agents.

\subsubsection*{\textbf{Positioning of \model}}

\model sits at the intersection of tabular data discovery, question answering, and emerging agentic search. Unlike conventional data discovery systems that rely on fixed retrieval pipelines and substantial offline indexing, \model performs adaptive discovery over large, heterogeneous tabular data lakes using an LLM agent. At the same time, unlike approaches based purely on direct corpus interaction, \model does not rely solely on unguided exploration of raw data. Instead, it combines lightweight, reusable semantic schema guidance with direct access to raw tables, allowing a small amount of offline structure to guide, rather than constrain, agentic search. Its thin harness exposes complementary table-specific primitives for semantic column search, value and schema inspection, SQL execution, and, in the join-based setting, discovering and verifying joinable columns, while leaving their composition and ordering to the agent. This enables \model to support both multi-table independent discovery and join-based discovery without assuming a fixed retrieval cardinality.

\section{Conclusions}\label{section:conclusions}

In this work, we extend tabular data discovery to multi-table settings in which relevant information may reside independently in multiple tables or require composition through joins. We introduce \model, a lightweight agentic framework that combines reusable column-level semantic guidance, structured tabular operations, general-purpose execution, and lightweight joinability metadata while leaving retrieval decisions to the LLM agent. We also construct a benchmark covering four independent-discovery datasets and two join-based datasets. Across all six datasets, \model achieves the highest table-discovery macro-F1 and the highest exact-match rate on both join-based benchmarks, while reducing offline preprocessing overhead relative to Pneuma and online latency and API cost relative to Claude Code. Its retrieval improvements also translate into higher downstream question-answering F1 when the same NL2SQL agent is applied to the discovered tables.



\bibliographystyle{ACM-Reference-Format}
\balance
\bibliography{sample-base}

\end{document}